# Topological Lattice Metamaterials – A Platform for Novel Electromagnetic Material Design Based on an Artificial Topological "Atom"

Wenjin Zhang,[1,2] Ziyuan Meng,[1] Zidong Zhang,[2,†] Ke Bi,[3,‡] Runhua Fan[4], Yi Du[1], Weichang Hao[1,*]

[1]*School of Physics, Beihang University, Beijing 100191, PR China*
[2]*School of Materials Science and Engineering, Shandong University, Jinan 250061, PR China*
[3]*School of Science, Beijing University of Posts and Telecommunications, Beijing 100876, PR China*
[4]*College of Ocean Science and Engineering, Shanghai Maritime University, Shanghai 201306, PR China*

Corresponding author.
*whao@buaa.edu.cn
†zhangzidong@sdu.edu.cn
‡bike@bupt.edu.cn

In nature, most materials are composed of atoms with periodic structures. Hence, it's impossible to introduce topological structures into their lattice compose, because the atoms as basic blocks cannot be modulated. However, the lattice compose of metamaterials can be designed conveniently. In our work, we propose to introduce topological non-trivial structures, Möbius unknots, as the basic block (the artificial chiral "atoms") to design metamaterials. A 5.95 GHz intrinsic peak, in addition to the electrical resonance peak near 11 GHz on the transmission coefficient spectrum was confirmed by theoretical calculations, finite-difference time-domain (FDTD) simulations and experiments when electromagnetic waves transfer to a chiral Möbius unknot. Theoretical analysis indicates that this intrinsic peak originates from the phase transition caused by the electromagnetic waves propagate along the Möbius unknot non-trivial structure. It is similar to the state of spin-splitting of electron levels. Take the artificial chiral "atoms" - Möbius unknots as the basic block, we can construct two-dimensional and even three-dimensional ordered metamaterials. The simulation and experimental results showed that the response to electromagnetic wave in the GHz band can be modulated by the coupling between the periodic potential and the spin-like of energy levels.

Topology in mathematics aims to identify and describe invariant properties of objects under continuous deformation. In last decade, topology has been successfully introduced into band theory in physics, in which all Bloch states can be labelled by momentum $k$ and can be constitute the vector bundle over entire first Brillouin zone. The topological properties of these vector bundles offer new insights in exotic electronic phases and phenonemona, including quantum Hall effect (QHE) [1,2], non-trivial states in topological insulators [3] and various semimetals. While it has achieved a huge success in identification of invariants of Bloch states in reciprocal lattice, the analogue form in real lattice remains unclear. This is because the fundamental building block of solid is the atom which does not possess dontinuous deformation in real space.

Metamaterials, the artificial composites that acquire exotic electromagnetic properties from assemblies of multiple artifical geometric elements, provide an excellent platform to achieve analogue of topological Bloch states in real space [4,5]. The continuous deformation of electromagnetic properties can be realized by controlling the geometric parameters of the elements in metamaterials as well as proper electromagentic waves chosen [6,7]. Examples include various metamaterials, such as split ring resonators (SRRs) [8,9], split-cube [10], paired rod [11,12], fishnet [13], etc.

Recently, some research groups introduced chiral structures into metamaterials, including helix [14, 15], chiral bridge [16], metal rosettes [17], helical wire [17] and so on. From differential geometric view, all these metamaterials are still classified as topological trivial structures [8, 9, 10, 11, 12, 13, 14, 15, 16, 17, 18, 19].

Similar to topological characteristics in Brillouin zone, topological non-trivial structures should be the fundamental lattice elements in metamaterials in order to introduce topological features. In this letter, we propose a new strategy to utilize Möbius unknot, the topological non-trivial structure, as the basic block to design topological lattice metamaterials (TLM). In our proposal, the topological non-trivial structure of Möbius unknot will lead to the accumulation of Berry's topological phase shift, then a chiral energy level splitting, which is confirmed by calculations from Master equation. These chiral energy levels provide new physics and possibilities to design novel electromagnetic (EM) metamaterials.

Aharonov–Bohm phenomena shows that phase shift could accumulate in the topological non-trivial structures. Similarly, there's a twist phase in the joining process of Möbius unknot, which doesn't exist in trivial ring. It results from the basic feature that the Möbius unknot and the trivial ring correspond to non-trivial U(1) bundle and trivial U(1) bundle, respectively [20, 21, 22, 23, 24]. Any vectors, which parallel to the surface of the Möbius unknot, turn around and head to the opposite direction after one revolution around the ring, which is similar to berry's topological phases in optical fiber [25, 26, 27]. This strange condition shows the overall rotation angle of the vector of state, representing the non-integrable phase factor of the wave function, which is also known as the geometric phase of vector. This informs us, the integral of magnetic field could equal to a non-zero integer for non-trivial bundle. In order to confirm our proposal, we establish models of trivial ring and Möbius unknot in Figs. 1(a) and 1(b), then calculate the energy levels of their eigenstates, the non-trivial states of Möbius unknot are similar to energy states of electron levels, as shown in Figs. 1(c) and 1(d).

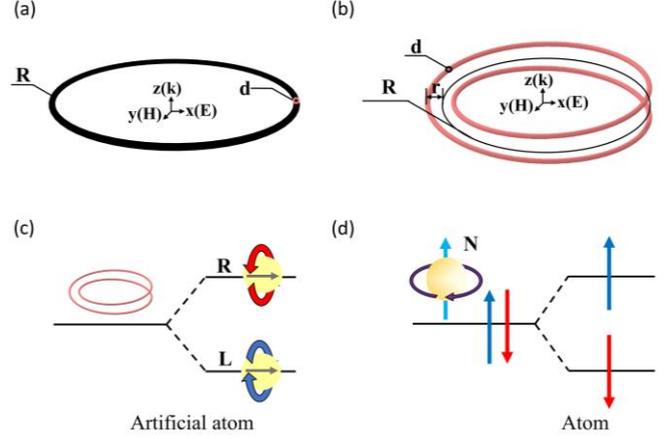

FIG. 1. Models of trivial ring and Möbius unknot. The structure parameters of (a) trivial ring and (b) Möbius unknot, $R = 4$ mm, $d = 0.2$ mm, $r = 0.5$ mm. Degenerate and splitting energy states of (c) an artificial atom (Möbius unknot) and (d) a normal atom, split by their chirality.

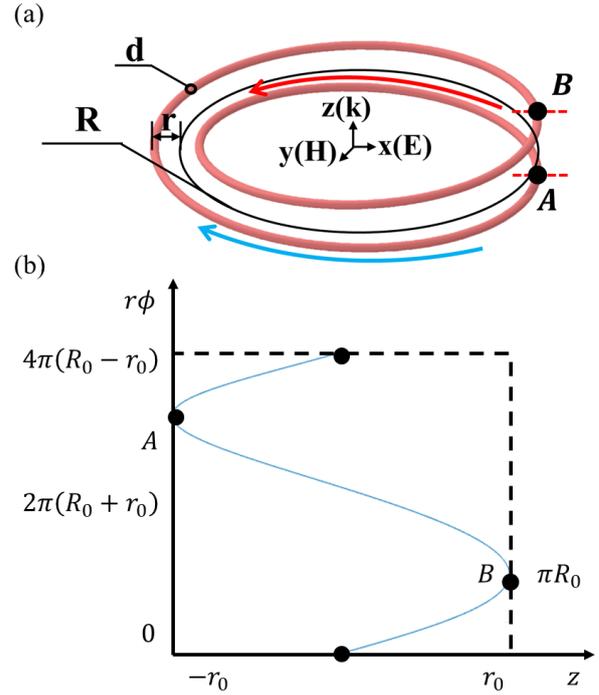

FIG. 2. (a) Model of Möbius unknot, made up of two helixes end to end. (b) The solid line represents the path of the Möbius unknot.

## Results
**Theory**. Theoretical derivation predicts a nonzero geometric phase shift,
$$\gamma(C) = \Omega(C_1) - \Omega(C_2) = 4\pi r(1/s_2 - 1/s_1), \quad (1)$$
Which could be accumulated by two electrons each travels $2\pi$ on the opposite directions along the unknot. Here, $C_1$ and $C_2$ are the paths of the electrons and they make up of the complete Möbius unknot, $s_1$ and $s_2$ are the lengths of the two paths, separately. The nonzero phase shift $\gamma(C)$ then results in the interference absorption of specific electromagnetic wave after one round.

**Master equation calculation.** For a Möbius unknot in Fig. 1(b), we consider the master equation for chiral objects in EM field for radiation of frequency $\omega$ in the Möbius unknot, [28]

$$\nabla^2 \boldsymbol{E}(\boldsymbol{r}) + 2\omega\mu\xi\nabla \times \boldsymbol{E}(\boldsymbol{r}) + K^2\boldsymbol{E}(\boldsymbol{r}) = 0, \quad (2)$$

Here, $\mu$ is the magnetic permeability, $\xi$ is the chirality admittance, and $\boldsymbol{E}(\boldsymbol{r})$ is the EM eigenmode. Combined with the path of Möbius unknot,

$$\boldsymbol{l}(\varphi) = (\rho, \varphi, z)$$
$$= \left(R + r\cos\frac{\varphi}{2}, \varphi, r\sin\frac{\varphi}{2}\right), 0 \leq \varphi \leq 4\pi, \quad (3)$$

Among, $R$ is the radius and $r$ is the offset. For the EM field $\boldsymbol{E}(\boldsymbol{r})$ along the unknot, $\boldsymbol{E}(\rho, \varphi, z) \propto d\boldsymbol{l}(\varphi)/d\varphi$.

Take the phase shift into account, the energy eigenvalues along this Möbius unknot are near,

$$E_\pm = hf_\pm = \frac{h\omega_\pm}{2\pi} \cong \frac{hc}{2\pi R} \pm \frac{1.934h}{4\pi\varepsilon R}|\xi|, \quad (4)$$

Here, for the intrinsic frequency $f_0$,

$$f_0 = \frac{1}{2\pi}(\omega_+ - \omega_-) \cong \frac{1.934}{2\pi\varepsilon R}|\xi|, \quad (5)$$

In the above functions, $C$ is a coefficient between 0.09 and 0.11 in linear fitting and $c$ is the velocity of light. The theory calcaulation showed there is an abnormal EM response of intrinsic peak of $1.934|\xi|/2\pi\varepsilon R$ resulted from the energy states divided by the chirality admittance $\xi$, which can be positive or negative depending on the handedness of chiral object. Meanwhile, the intrinsic frequency doesn't exist in trivial ring, since chirality admittance is zero so that $E_{n+}$ strictly equals to $E_{n-}$. Here, we assume that $R = 4 \text{ mm}$, $d = 0.2 \text{ mm}$, $r = 0.5 \text{ mm}$. Based on our theoretical result, we predict a special characteristic frequency near 6 GHz due to the non-integrable phase factor caused by the topological structure of Möbius unknot.

**Finite-difference time-domain.** In order to check our theoretical calculation result, we apply finite-difference time-domain (FDTD) to simulate the response of Möbius unknots in EM field. We use copper, aluminum and silver as the objects' materials for simulation, Figs. 3(a) and 3(b) are transmission coefficient spectrums of trivial ring and Möbius unknot. Comparing with the trivial ring, we can find that there is a special peak at 5.95 GHz in the spectrum of Möbius unknot besides the 1st order electrical resonance peak near 11 GHz [29, 30], which is consistent with our theoretical calculation. The electrical resonance peak at 11 GHz in transmission coefficient spectrum of both Möbius unknot and the trivial ring resulted from Mie resonance when the frenquency of EM wave equals to the inherent frequency of basic block [30, 31, 32]. The Mie resonance in our artificial 'atoms' corresponds to the electron orbital levels in atoms. As shown in Fig. 3(c), we increase the radius $R$ of the Möbius unknot to investigate the change of intrinsic peak. It is obvious that, the position of the intrinsic peak moves toward the lower frequency as the radius increases, and a perfect linear relationship between the frequency $f$ and reciprocal of radius $1/R$ is shown in Fig. 3(d), which is consistent with the prediction result of Equation (4). In addition, inset in Fig. 3(c) shows the variation of the integral area with the $R$. As the volume of the unknot increases, the energy loss of electromagnetic waves will also increase, so the peak intensity and integral area increase in the transmission spectrum. We monitored the surface current at the position of Möbius unknot intrinsic peak and the first-order electrical resonance peak in Fig. 3(e),(f). Unlike the surface current of electrical resonance peak, the direction of the Möbius unknot's surface current at 5.95 GHz is not affected by the electric field component. The direction of its surface current is even opposite to that of the electric field. This demonstrates that there are two kinds of 'electrons' states in the artificial atom, which are converted from the degenerate state to the two states under the excitation of the electromagnetic wave, and changing the incident direction of the electromagnetic wave does not affect the intrinsic peak in Fig. S6. The robustness of intrinsic peak results from topological feature of basic block [33].

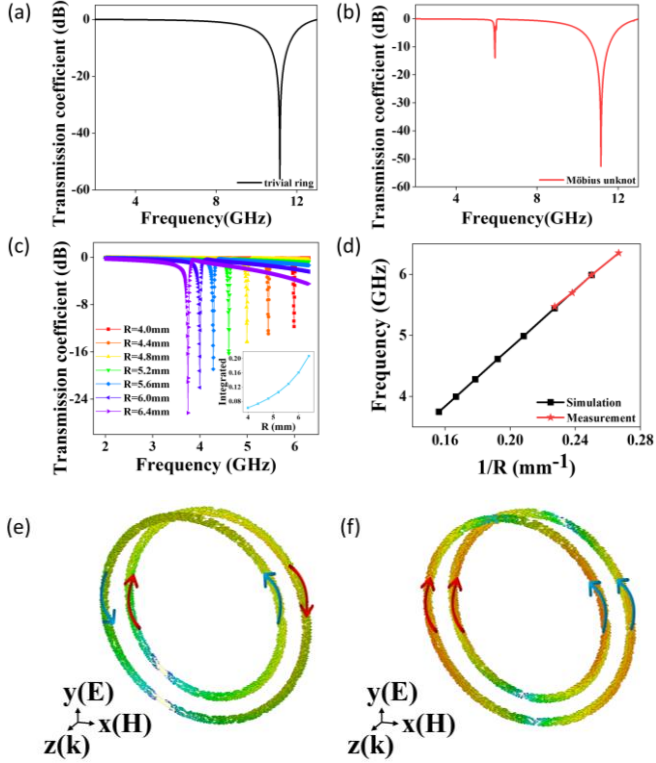

FIG. 3. Transmission coefficient spectrum of (a) the trivial ring and (b) Möbius unknot. (c) The transmission coefficient of the Möbius unknot varies with the radius R, and the inset is integral area of the intrinsic peaks vs scale factor. (d) The frequency of the intrinsic peak varies with the reciprocal of the radius $1/R$. (e) Surface current on the Möbius unknot at 5.95 GHz. (f) Surface current on the Möbius unknot at 11 GHz.

It's worth noting that the absorbing of electromagnetic wave at the intrinsic peak is similar to the spin-splitting state shown in Fig. 1.
Figs. 4(a) to (d) show the bistatic scattering RCS of trivial ring and Möbius unknot. The RCS of Möbius unknot in left-handed circular polarization field (Figs. 4(b)) and right-handed circular polarization field (Figs. 4(d)) have different scattering intensities by using a negative-z direction propagating plane-wave (frequency = 5.95 GHz) as excitation source, while the RCS of a trivial ring is exactly the same in different circular polarization field as shown in Figs. 4(a) and 4(c). Their response to electromagnetic waves in different polarization fields indicates that when the base space is non-trivial, the RCS should be determined by the topological feature of the basic block, as shown in the Figs. S7(d)-(f). Interestingly, under the excitation of a circular polarization plane wave, the surface currents of Möbius unknot rotate with the direction of rotation of the electric field component as the phase changes in Fig. S8, therefore, the surface current's period of change is $\pi$ as shown in the Fig. S9. Whereas, under the excitation of the linear polarization plane wave, the surface current will rotate clockwise, which is also the strong evidence of the Möbius unknot's topological feature.

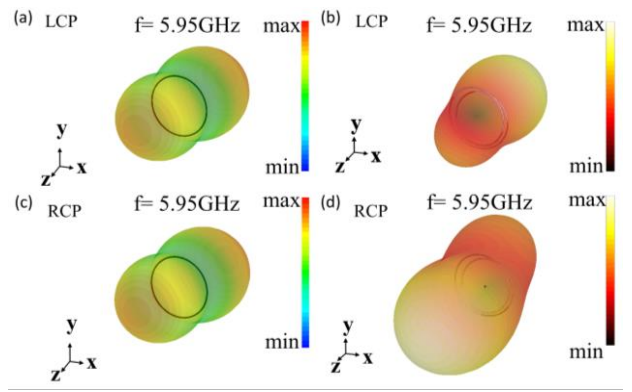

FIG. 4. The bistatic scattering RCS of (a),(c) trivial ring and (b),(d) Möbius unknot by using plane-wave simulation under different circular polarization mode, (a),(b) correspond to LCP, (c),(d) correspond to RCP.

**Free space method in anechoic chamber.** To further confirm our theoretical calculation and simulation results, the EM response of trivial ring and Möbius unknot made of copper was characterized by free space method in anechoic chamber. The observed transmission coefficient spectrum are shown in Fig. 5. The simulation results (Figs. 5(a),(c)) are in good agreement with experimental data (Figs. 5(b),(d)). Fig. 5(b) illustrates the response of the copper trivial ring sample to electromagnetic waves, as our theory predicts, an electrical resonance peak appears near 11 GHz, which is consistent with the simulation results in Fig. 5(a) on the left. The peak intensity of the resonant peak of the measurement results is lower than that of the simulation datas.

This varience is because of the conductivity difference between the real copper material and the ideal one, so the intensity of surface current in test sample decreases, reducing the electrical resonance peak's strength. Fig. 5(d) shows the copper Möbius unknot's response to EM waves. As predicted by theoretical calculation, in addition to the electrical resonance peak, a slightly weaker intrinsic peak appears at 5.95 GHz. Just like the result of trivial ring, the peak intensity in the measurement is weaker than that in the simulation results in Fig. 5(c). As shown in Figure S12, the type of materials selected for making the ring hardly affects its transmission coefficient spectrum.

As shown in Fig. 6(a), the Möbius unknot is arranged in a square lattice form, which we call 2D topological lattice metamaterial (TLM). The unit distance $a$, polarization direction of electromagnetic waves and the Brillouin zone and irreducible Brillouin zone of the square lattice are marked, respectively. Fig. 6(b) is 2D TLM sample made of copper. Fig. 6(c) shows the band diagram of a TLM. We found that there is a photonic band gap around 5.95 GHz, which is completely consistent with our theoretical calculations. In the anechoic chamber, the transmission coefficient spectrum obtained by the free space method is shown in Fig. 6(d), the peak position in the measurement data is exactly consistent with the simulation results. The transmission coefficient spectra of two-dimensional Möbius metamaterials with different lattice constants are numerically calculated, shown in Fig. 6(e). As the lattice spacing increases, the leakage of electromagnetic energy increases, so the integral area of intrinsic peaks in the transmission spectrum decreases, as shown in Fig. 6(f). Fig. 6(g) shows that the positions of electrical resonance and intrinsic peak change with the distance between the units. The position of the electrical resonance peak fluctuates greatly, but the position of the intrinsic peak is almost always near 5.95 GHz. This indicates that the intrinsic peak is the internal property of the Möbius unknot, which cannot be changed by the arrangement forms. RCP wave and LCP wave were respectively used to excite Möbius metamaterial, and the transmittance spectrum was obtained in Fig. 6(h). It can be found that the response of metamaterial under the two polarization modes is different at 5.95 GHz, and the peak strength corresponding to LCP is stronger, which means that the energy loss of the topological metamaterial to LCP wave is larger than that of RCP wave because the topological structure has internal chiral properties. For square arrangement, there are five common two-dimensional crystal structures including the energy loss of the topological metamaterial to LCP wave is larger than that of RCP wave because the topological structure has internal chiral properties. For square arrangement, there are five common two-dimensional crystal structures including hexagonal, centered rectangular, oblique, honeycomb and rectangular. We calculated the transmission spectrum of the two-dimensional metamaterials arranged in all forms by Möbius unknot with different unit distance. The positions of electrical resonance peaks and intrinsic peaks are shown in Figs. S10. The shift of peak position in all five different lattice metamaterials have the same tendency, as well as the integral areas with similar pattern shown in Fig. S11. So, a class of two-dimensional metamaterials can be designed using the Möbius unknot internal property, which is minimally affected by the lattice spacing, compared to the electrical resonance frequency.

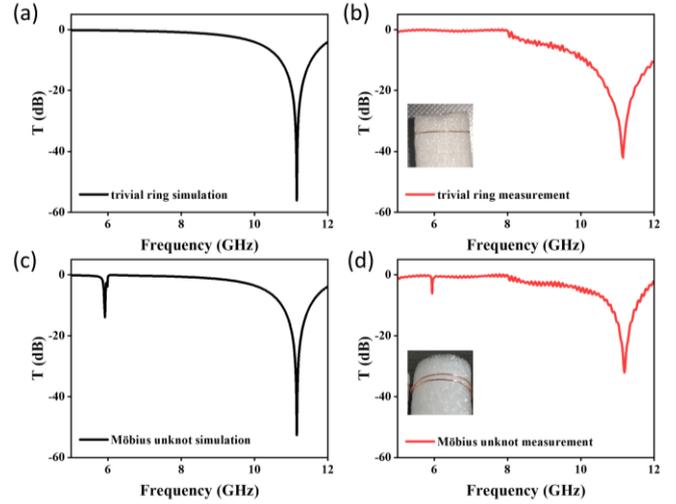

FIG. 5. The transmission coefficient spectrum of copper trivial ring is obtained by (a) simulation and (b) measurement, and the transmission coefficient spectrum of copper Möbius unknot is obtained by (c) simulation and (d) measurement.

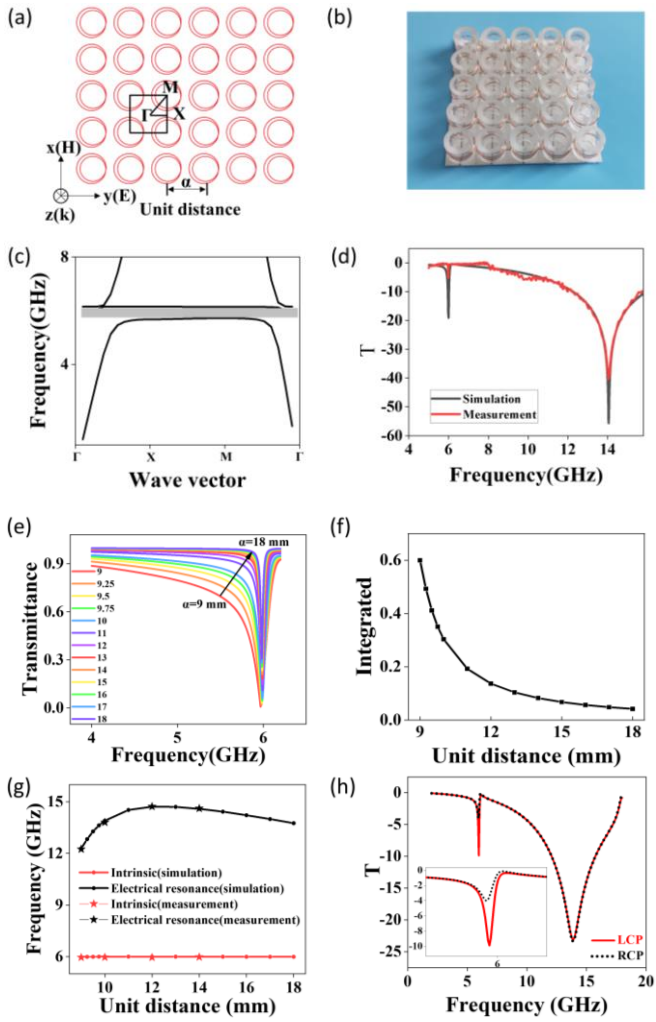

FIG. 6. 2D Möbius metamaterial (a) The Möbius unknot is arranged in the form of two-dimensional squared crystal, Brillouin zone and the irreducible Brillouin zone are marked. (b) 2D Möbius metamaterial sample made of copper. (c) Band gap diagram of two-dimensional squared metamaterials. (d) Experimental observation of transmission coefficients of 2D topological lattice metamaterial sample made of copper, black line corresponds to simulation and red line corresponds to measurement, the distance between the two units is 10 mm. (e) The transmission coefficient of the 2D topological lattice metamaterial varies with the unit distance $a$. (f) Integral area of the intrinsic peak vs the unit distance. (g) The position of electrical resonance peak (black lines) and intrinsic peak (red lines) vs the unit distance, the points marked by the pentagram correspond to measurement data. (h) Transmission coefficient spectra of two-dimensional Möbius metamaterials under the excitation of different circularly polarized waves.

## Discussion

In summary, we design and fabricate an artificial metamaterial composed of periodic array of topological non-trivial structures, resulting topological non-trivial Bloch states, which split from a single trivial Bloch state. The energy gap between the two unique non-trivial states could only be determined by the perimeter of the chiral array unit in real space, as the intrinsic frequency is inversely proportional to the radius, corresponding to the Brillouin zone determined by lattice structure for trivial topological Bloch states in reciprocal space. We realized the adjustment of non-trivial states by changing geometric parameters of array units in real space, despite of their arrangements, which could be utilized to manufacture stable devices, including antennas, filters, and resonators, with minimal interference of neighboring units.

## Acknowledgements

The authors thank Dr. Wei Li, Dr. Yin Jiang, Dr. Xin Liu and Prof. Shoushu Gong for stimulating discussions. This work was financially supported by Beijing Natural Science Foundation (Z180007) and National Natural Science Foundation of China (Grant Nos. 11874003, 51672018, 51601105).

## Author contributions
W. Z. and Z. M. contribute equally to this work.

## Competing interests
The authors declare no competing interests.